\documentclass[12pt,preprint]{aastex}
  
\newcommand{\el}{SN~1999el}  
\newcommand{\kms}{km s$^{-1}$}

\shorttitle{Supernova SN 1999el}
\shortauthors{Di Carlo et al.}

\begin{document}
\title{Optical and infrared observations of the supernova SN 1999el\footnote{Based on observations  
 collected with the AZT-24 telescope (Campo Imperatore, Italy), 
 the 1.8-m telescope of the Osservatorio 
 Astronomico di Padova (Asiago, Italy), the TNT telescope operated by the Osservatorio 
 Astronomico di 
 Collurania-Teramo (Teramo, Italy), the Italian National Telescope TNG (La Palma, Canary Islands, Spain), 
 the 1-m telescope of the Observatorio Astron\'omico Nacional (Tonantzintla, Mexico), 
 the 2.1-m telescope of the Guillermo Haro Observatory (Cananea, Mexico).}}  
\author{E. Di Carlo\altaffilmark{2}, 
F. Massi\altaffilmark{2}, 
G. Valentini\altaffilmark{2}, A. Di Paola\altaffilmark{3},  
F. D'Alessio\altaffilmark{3}, \\ E. Brocato\altaffilmark{2},
D. Guidubaldi\altaffilmark{2}, M. Dolci\altaffilmark{2}, 
F. Pedichini\altaffilmark{3}, R. Speziali\altaffilmark{3},\\  
G. Li Causi\altaffilmark{3}, A. Caratti o Garatti\altaffilmark{2},
E. Cappellaro\altaffilmark{5}, M. Turatto\altaffilmark{5}, 
A. A. Arkharov\altaffilmark{4}, \\ Y. Gnedin\altaffilmark{4},
V. M. Larionov\altaffilmark{4,10,11},
S. Benetti\altaffilmark{5}, A. Pastorello\altaffilmark{5},
I. Aretxaga\altaffilmark{8}, \\ V. Chavushyan\altaffilmark{8}, 
O. Vega\altaffilmark{8}, I. J. Danziger\altaffilmark{9}, 
A. Tornamb\'e\altaffilmark{2}}
\email{dicarlo@te.astro.it}
\altaffiltext{2}{Osservatorio Astronomico di Collurania-Teramo, Via M. Maggini, 
I-64100 Teramo, Italy}
\altaffiltext{3}{Osservatorio Astronomico di Roma, Via Frascati 33, 
I-00040 Monteporzio Catone (Roma), Italy} 
\altaffiltext{4}{Central Astronomical Observatory at Pulkovo, Pulkovskoe shosse 65,
196140 Saint-Petersburg, Russia}
\altaffiltext{5}{Osservatorio Astronomico di Padova, 
Vicolo dell' Osservatorio 5, I-35122 Padova, Italy}
\altaffiltext{6}{Istituto Nacional de Astrof\'isica, \'Optica y Electr\'onica, 
Apt.do Postal 51 y 216, Puebla, Mexico}
\altaffiltext{7}{Telescopio Nazionale ``Galileo'', Apartado de Correos 565, 
E-38700, Santa Cruz de La Palma, Canary Islands, Spain}
\altaffiltext{8}{Istituto Nacional de Astrofisica, Optica y Electronica INAOE, 
S.M. de Tonantzintla, Puebla, Mexico}
\altaffiltext{9}{Osservatorio Astronomico di Trieste, Via G.B. Tiepolo 11, 
I-34131 Trieste Italy}  
\altaffiltext{10}{Astronomical Institute of St. Petersburg University, Russia}
\altaffiltext{11}{Isaac Newton Institute of Chile, St. Petersburg branch}
\label{firstpage}  
  
\begin{abstract}     
Optical and near-infrared light curves of the Type IIn  supernova 1999el   
in  NGC 6951 are presented. A period of   
220 days (416 days in the near-infrared)
is covered from the first observation obtained a few days before  
maximum light.  
Spectroscopic observations are also discussed.   
Using as a distance calibrator the Type Ia SN 2000E, which occurred some months
later in the same galaxy, and fitting a blackbody law to the photometric data
we obtain a maximum bolometric luminosity for SN~1999el of $\sim 10^{44}$ erg s$^{-1}$. 
In general, the photometric properties of SN~1999el are very  
similar to those of SN~1998S, a bright and well studied Type IIn SN, 
showing a fast decline in all observed bands  
similar to those of  Type II-L SNe. The differences with SN~1998S 
are analyzed and ascribed to the differences in 
a pre-existing circumstellar envelope in which dust was already present at the 
moment of the SN outburst. We infer that light echoes may play a possibly 
significant role in affecting the observed properties of the light curves, 
although improved theoretical models are needed 
to account for the data.
We conclude that mass loss in the progenitor RG stars is episodic and
occurs in an asymmetric way. This implies that collapsing massive stars 
appear as normal Type II SN if this occurs far from major mass loss episodes,
whereas they appear as Type IIn SNe if a large mass loss episode is in progress.  
  
\end{abstract}       

  
\section{Introduction}     
                       
A significant fraction of the total number of Type II supernovae (SN~II) show 
distinctive features which led Schlegel (1990) to introduce the class of 
SN~IIn. These objects are characterized by a blue continuum near maximum and a slow 
temporal evolution of the spectrum; the slowly declining luminosity in the 
optical bands and the narrow ${\rm H}{\alpha}$ line profile superimposed on 
a broad base indicate the presence of diffuse lower velocity matter around the site of
the explosion. Because of their relatively high luminosity and slowly declining light 
curve, the SN~IIn are easier to discover than faint, narrow peak light curve 
objects. Their intrinsic frequency has been estimated by Cappellaro et al. 
(1997) to be 5\% of all SN~II. There is, however, a marked inhomogeneity among the 
events (Cappellaro \& Turatto 2001), namely concerning the decline rate of the light 
curves, a matter which merits understanding. 
It has not yet been clarified to what extent individual differences among members 
of this type of SNe can be totally ascribed to the physical properties
of pre-existing circumstellar matter (CSM).
 
In fact, recent attempts to explain the long term   
properties of visual light curves (LCs)  with light echoes 
from circumstellar matter
by Roscherr \& Schaefer (2000) have not been fully successful.   
There are observational 
features which disagree with echo model predictions, forcing  
these authors to conclude that any 
pre-existing dust around the site of the explosion must be  
almost completely destroyed at the moment of the outburst.   
However, this contrasts with the observed late infrared 
emission from SN~1998S, which has been inferred to result from 
dust cooling (Fassia et al. 2000). Based on their observational data,  
Fassia et al.\ argue that dust was already present within the  
precursor wind of SN~1998S, and hence at least
a fraction of it must have survived the explosion. 
This suggests that improved observations are needed in order to 
understand the environment of SN sites from this
diagnostic tool.
So far only a few of the models developed to test the effects of light echoes
have taken into account the near infrared (NIR) light curves
(see Emmering \& Chevalier 1988).  
 
On the other hand, the slow decline of light curves for Type IIn SNe appears to be well   
accounted for by simple models of ejecta-wind interactions   
(Terlevich 1994; Plewa 1995).  
At the same time, the peculiar shape of the emission lines has been 
also attributed  to ejecta-wind interactions involving a two-component 
wind, characterized by  dense clumps (or a dense equatorial wind) associated 
with a less dense spherically symmetric distribution (Chugai \& Danziger 1994).  
 
Therefore, several open questions remain, which can be formulated 
as follows:  
\begin{enumerate} 
\item How important is the issue of the  
inhomogeneity of this class of events and  
what causes this inhomogeneity?    
\item What is the evolutionary scenario causing some  
massive precursors of SNe 
to explode within a pre-existing envelope 
of diffuse circumstellar matter and what physical mechanisms 
affect the observational properties?  
\item Is dust present in the CSM before the explosion? 
\item To what extent do echo processes and  
ejecta-wind interactions determine the observational properties of such SNe? 
\end{enumerate}

In this paper we present optical and infrared data for the Type IIn 
SN 1999el. The present work provides new photometric
measurements and improves the
observational constraints on this class of SNe. 
Observations and data 
reduction are presented in section 2. The results are presented and discussed 
in sections 3 and 4, and summarized in section 5.

\section{Observations and data reduction}  
  
%
%
\begin{figure}  
\vskip 3cm  
\caption{$V$ band image of SN~1999el in the galaxy NGC~6951 obtained at
 the TNG on JD 2451690. Stars of the local sequence are indicated by
 numbers (\#6 is out of the displayed field). The enlarged box shows the supernova, which lies towards a
 group of close projected stars referred to as c1, c2 and c3 in the text.
\label{gal:sn}}  
\end{figure}
SN~1999el was discovered on October 20.45 UT  
(julian day 2451472) at mag $\sim 15.4$ (unfiltered)
by the BAO SN Survey (Cao et al.\ 1999, IAUC 7288)
$\sim 23.4"$ east and $\sim 9"$ south of the galactic
nucleus of NGC~6951, an inclined  
($i=42 \degr$) barred spiral galaxy hosting a type 2 Seyfert AGN, 
located at $\alpha(1950)=20:36:36.59$ and $\delta(1950)=+65:55:46.00$
(see Fig.~\ref{gal:sn}). For a list of galactic properties
see, e.\ g., Kohno, Kawabe, \& Vila-Vilar\'o (1999) and references therein.
The distance to NGC~6951 was recently
discussed by Vink\'o et al.\ (2001). The estimated distance moduli range from
31.85 mag (23.4 Mpc) to 32.59 mag (33 Mpc), where the lowest values are based on the
Tully-Fisher relation and the highest ones on the multi-color light curve shape (MLCS) method applied to the
SN~2000E (a Type Ia SN). We adopt our own determination ($32.1 \pm 0.4$ mag) based on
our observations of SN~2000E (see Valentini et al.\ 2002, in preparation). 
 
Cao et al.\ (1999) report no trace of the SN 
on images taken on October 19.2 by KAIT (to a limiting magnitude of 18) and on 
October 19.44 by BAO, so we can assume October 20.00 as the
zero point for the light curves, since it must be very close 
(within a few hours) to the shock breakout time. 
According to Cao et al.\ (1999), an image taken on October 21.44 
indicates a magnitude of $15.0$ for the supernova, whereas on October 22.2 
it appeared at a magnitude of $15.2$ (no band is indicated). 
It can be seen from Fig.~\ref{gal:sn} that
SN~1999el is projected between two nearby foreground stars (c2 and c3)
with a separation of no more than a couple of arcsec. 
 
\subsection{Optical data: acquisition, reduction and photometry}  
  
Optical photometry of \el\ has been obtained on images collected   
with several telescopes at different sites in the northern hemisphere.    
These were reduced using the standard techniques for de-biasing and 
flat-fielding. 

One set of data comes  
from the 0.72-m  Teramo-Normale Telescope (TNT), at the  
Teramo Observatory (Italy).  The TNT is an f/14 Ritchey-Chretien  
reflector  equipped with a Tektronic Tk512 CB1-1 front-illuminated  
$512 \times 512$ pixels CCD.  The pixel size of 27 $\mu$m yields a scale of  
$0.46$ arcsec/pixel.  During each observational night we obtained 4 to 12  
frames per filter in the B, R and I bands.
  
Other photometry was performed on images
obtained with the 1-m telescope operated by the 
Observatorio Astron\'omico Nacional in Tonantzintla (Mexico), 
equipped in the f/15 Cassegrain port with a 1024$\times$1024 Thomson 
THX~31156 CCD and a Cousins $R$ filter. The field of view covered by this 
configuration is $4.2'\times4.2'$. The chip was binned $2\times2$ and read 
in standard mode (3.47 $e^-$/pixel, 2.7$e^-$/ADU). The effective on-chip 
integration of the SN field was split into 3 to 6 on-chip integrations of 
300 s, offset from each-other by $\sim 20''$. 

For imaging, also the 1.8-m telescope of the Osservatorio Astronomico di 
Padova  sited in Asiago, Cima Ekar (Italy), equipped with AFOSC, was used. The 
instrument is a focal reducer type spectrograph/camera with a 1K 
$\times$ 1K Site Thinned CCD ($24\mu$m). The scale of $0.473 \arcsec$/pixel 
gives a field of $8 \times 8$ arcmin$^{2}$. 

Some  photometric data were acquired
using the 2.1-m telescope of the Guillermo Haro Observatory in 
Cananea (Mexico), equipped with the spectrophotometer LFOSC at the f/12 
Cassegrain focus. Johnson $V$ and Cousins $R$ filters were used. The chip 
is a $385\times578$ EEV~P8693, with a field of view of $6'\times10'$. This 
was read in the standard readout mode: 8$e^-$/pixel, 4.7$e^-$/ADU. The 
effective on-chip integration of the SN field was split into 2 to 3 
integrations of 60 s in the $R$-band and of 120 s in the $V$-band. 
 
Late time $UBVRI$ photometry was obtained with the Optical Imager Galileo 
(OIG) mounted on the 3.58-m Italian National Telescope (TNG, Roque de los 
Muchachos, La Palma, Canary Islands) and equipped with a mosaic of two 
thinned, back-illuminated EEV42-80 CCDs with $2048\times 4096$ pixels each (pixel 
size $13.5$ $\mu$m; pixel scale in $2\times2$ binned mode $0.144 \arcsec$/pixel). 
These nights were photometric and were used to calibrate the 
local sequence around the SN (1 to 8, see Fig.~\ref{gal:sn}) 
which in turn was used for calibrating non-photometric nights. 
The instrumental color equations were obtained through observations of 
stars in the standard fields of Landolt (1992).
Three of the local standards (1, 3 and 5) are a subset of 
those selected by Vink\'o et al. (2001) to carry out differential
photometry ($BVRI$) on SN2000E (their F1, B1, B3, respectively).
The magnitudes adopted in our and their study for those sources 
agree within $0.05$ mag, with two exceptions: we
found our star 1 to be $0.13$ mag brighter in $R$ and our star 5 to be
$0.13$ mag brighter in $I$. The agreement can be considered good and confirms
the validity of both our calibration and that of those authors.
However, our measurements were obtained using a larger telescope
(the TNG) under very good seeing and photometric conditions. This
should have afforded us a better sky subtraction which, particularly at longer
wavelengths, may be easily affected by nebular emission from the host galaxy. 

During the \el\ follow-up observations,
another SN event occurred within NGC~6951 (i.\ 
e., SN~2000E), so the CCD at the TNT was operated to 
image the 2 objects on the same 
frame. The exposure times were chosen so as to achieve the best 
signal-to-noise ratio for both supernovae, but avoiding 
saturation of the brightest calibration star in the field.
 
Since both supernovae are well embedded in 
the galaxy background, we obtained instrumental magnitudes by point-spread 
function (PSF) fitting using the DAOPHOT package in IRAF or a set of 
procedures developed for SN photometry in the same environment. For the 
TNT and Tonantzintla observations, the final instrumental magnitudes were 
obtained as an average of all valid values derived for a given night.
 
Errors have been obtained with artificial star experiments, i.e. placing 
stars of the same magnitude as the supernova in parts of the galaxy 
arm characterized by similar complexity as the SN site, then computing the 
r.m.s. of the derived magnitudes.
For TNT images however, the uncertainty in the photometry has been
estimated by evaluating the PSF fit residuals and the r.m.s. of the 
$n$ successive measurements obtained in a single night for the SN 
after calibration through the local standards.
It is more difficult to assess the error 
due to the convolution of the PSFs of the SN and the nearby sources when
their brightness becomes comparable and the seeing is poor, 
which is the case 
in the TNT photometry at late epochs. Hence, we omit measurements 
which appear uncertain because of the inferior quality of the images.
Table~\ref{Table1} lists the results of optical photometry for \el.

\begin{deluxetable}{ccccccccccccc} 
\tabletypesize{\scriptsize}
\rotate
\tablecaption{$UBVRI$ photometry for SN~1999el obtained with different 
         instruments.
         \label{Table1}}       
\tablewidth{0pt}
\tablehead{
\colhead{JD} & \colhead{Epoch} &\colhead{$U$} & \colhead{$\Delta U$} & \colhead{$B$} & 
\colhead{$\Delta B$} & \colhead{$V$} & \colhead{$\Delta V$} & \colhead{$R$} & 
\colhead{$\Delta R$} & \colhead{$I$} & \colhead{$\Delta I$} & 
\colhead{Telescope} \\ 
 \colhead{(2451000+)}& \colhead{(days)} & \colhead{(mag)} & \colhead{(mag)} & \colhead{(mag)} & 
\colhead{(mag)} & \colhead{(mag)} & \colhead{(mag)} & \colhead{(mag)} & 
\colhead{(mag)} & \colhead{(mag)} & \colhead{(mag)}&  \colhead{}
}
\startdata
479.42 & 7.42 &  -   &  - & 15.87  &  0.06& -   & -  & 14.71   & 0.04&14.35&0.03&  TNT  \\ 
480.45 & 8.45 & -    & -  & 15.87 &  0.03& -   & -  & 14.68   & 0.04&14.31&0.02&  TNT \\ 
481.58 & 9.58 & -    &  - &     - &   -  & -   & -  & 14.69  & 0.02& -   & -  &  Tonantzintla \\ 
485.31 & 13.31 & -    & -  & 15.83 &  0.03& -   & -  & 14.60   & 0.03&14.22&0.03&  TNT \\ 
486.25 & 14.25 & -    &  - & 15.82 &  0.06& -   & -  & 14.56   & 0.06&14.25&0.02&  TNT \\ 
486.50  & 14.50 &15.70 &0.05& 15.73 &  0.05&15.11&0.04& 14.65  & 0.03&14.19&0.05&  TNG \\ 
486.52 & 14.52 &   -  &  - &   -   &   -  &-    & -  & 14.59  & 0.02& -   & -  &  Tonantzintla\\ 
486.57 & 14.57 &  -   & -  &    -  &   -  &15.04&0.03& 14.59  & 0.20& -   & -  &  Cananea\\ 
490.48 & 18.48 &15.65 &0.03& 15.64 &  0.06&14.98&0.02& 14.52  & 0.04&13.99&0.03&  Asiago\\ 
490.57 & 18.57 &  -   & -  &    -  &   -  &14.92&0.02& 14.48  & 0.04& -   & -  &  Cananea\\ 
493.60 & 21.57 &  -   & -  &   -   &  -   & -   & -  & 14.55  & 0.04& -   & -  &  Tonantzintla\\ 
498.68 & 26.68 & -    &-   &  -    &     -& -   & -  & 14.55  & 0.02&-    & -  &  Tonantzintla\\ 
506.21 & 34.21 &  -   & -  & 16.41 &  0.06& -   & -  & 14.86   & 0.06& 14.46 & 0.03 &  TNT \\ 
513.59 & 41.59 &  -   & -  &    -  &   -  & -   & -  & 15.08  & 0.10& -   & -  &  Tonantzintla\\ 
514.25 & 42.25 &  -   & -  & 16.76 &  0.06& -   & -  & 15.20   & 0.04& -   & -  &  TNT \\ 
516.53 & 44.53 &  -   & -  &    -  &  -   & -   & -  & 15.13  & 0.02& -   & -  &  Tonantzintla\\ 
520.21 & 48.21 &   -  & -  & 17.14 &  0.04&  -  & -  & 15.51   & 0.08 & 14.85 & 0.02 &  TNT \\ 
521.23 & 49.23 &  -   &  - & 17.17 &  0.06& -   & -  & 15.58   & 0.06&15.01&0.03&  TNT \\ 
522.54 & 50.54 &  -   & -  &    -  &    - & -   & -  & 15.33  & 0.03& -   & -  &  Tonantzintla \\ 
531.29 & 59.29 &  -   & -  & 17.52 &  0.08& -   & -  & 16.03   & 0.10& -   & -  &  TNT \\ 
561.25 & 89.25 & -    & -  & 19.44 &  0.30&  -  & -  & 17.54   & 0.20& -   & -  &  TNT \\ 
570.27 & 98.27 &   -  &   -& 19.67 &  0.20&  -  & -  & 18.40   & 0.20&17.54&0.07&  TNT \\ 
571.25 & 99.25 &  -   &  - &    -  &    - & 18.94&0.26& 18.46  & 0.14& -   & -  &  Asiago\\ 
572.24 & 100.24 &  -   & -  & 19.83 &  0.20&  -  & -  & 18.69   & 0.20&17.64&0.08&  TNT \\ 
575.29 & 103.24 &   -  & -  & 20.19 &  0.20&19.19&0.25& 18.25  & 0.18&17.53&0.22&  Asiago\\ 
585.30 & 113.30 &    - & -  & 20.27 &  0.16&19.22&0.16& 18.42  & 0.24&17.78&0.16&  Asiago\\ 
604.50  & 132.50 &21.27 &0.40& 20.97 &  0.30&19.52&0.25& 18.89  & 0.15&18.03&0.10&  TNG \\ 
605.50  & 133.50 &  -   &-   &   -   &   -  & $>19.17$ & -  & 19.27  & 0.30& -   & -  &  TNG \\ 
690.50  & 218.50 &21.44 &0.20& 21.54 &  0.15&20.49&0.10& 19.83  & 0.10&18.91&0.15&  TNG \\ 
698.54 & 226.54 &   -  & -  & 21.44 &  0.40&20.45&0.30& 19.81  & 0.30&18.90&0.25&  Asiago \\ 
699.55 & 227.55 &   -  &-   & 21.42 &  0.30&20.56&0.30& 19.77  & 0.30&  -  & -  &  Asiago \\ 
\enddata
\end{deluxetable} 
 

\subsection{Infrared data: acquisition, reduction and photometry}

Most near-infrared (NIR) observations of \el\ were obtained   
at the AZT-24 1.1-m telescope in Campo Imperatore  
(Italy) with SWIRCAM, during a period   
spanning October 25, 1999 to May 26, 2000.   
SWIRCAM (D'Alessio et al. 2000) incorporates a $256 \times 256$ HgCdTe NICMOS-3(-class) 
detector which, at the focus of AZT-24, yields 
a scale of $ 1.04 \arcsec$/pixel, resulting in a  
field of view of $\sim 4 \times 4$ arcmin$^2$.  The observations were  
performed through standard $J$ (1.25 $\mu$m), $H$ (1.65 $\mu$m) and $K$  
(2.20 $\mu$m) broad-band filters.  
Any on-source image was obtained as a median composition of five   
frames with a 15 arcsec dithering and integration times of 60 s each   
at the $J$ and $H$ bands, and 120  
s each at the $K$ band.  Off-source frames were taken with a 10 arcmin offset with the same  
observational procedure and exposure times, but a larger (50 arcsec)  
dithering. Sky images obtained as a median of the original  
dithered off-source frames were subtracted from the single on-source   
images before composition.  

Flat-field frames were acquired at twilight using the  
differential flat technique that relies on the natural variation of  
the sky background. In this way, the frames with the highest mean counts are 
combined together (through median filtering) in one single image, as are those with the lowest mean counts, and the resultant
lower signal image is subtracted from the resultant higher signal one  
so as to remove any biases and dark  
current.  This is used to flat-field each image after sky-subtraction.  
  
Deep $JHK_{\rm s}$ images of \el\ were obtained at the
3.58-m TNG telescope (La Palma)
with ARNICA during the night of May 26, 2000. ARNICA is a $256 \times  
256$ pixels NIR camera (Lisi et al.\ 1996) and was matched to the  
TNG with a $0.35$ arcsec/pixel scale (yielding a $90" \times 90"$ field
of view). Sequences of 4 dithered on-source and 4 off-source images were taken
moving the telescope to a sky position after each on-source integration.
Total on-source integration times amounted to 
12 minutes at $K_{\rm s}$ and 8 minutes at $J$ and $H$. 
Five images per band of a group
of standard stars (AS 40--5, AS 40--0 and AS 40--1; Hunt et al.\ 1998)
were also obtained so as to locate the brightest of them (AS 40--5)
at the center of the array and of each detector quadrant in turn.
Flat-field frames in the $K_{\rm s}$ band were taken at sunset.
Only the $K_{\rm s}$ frames were sky-subtracted after
flat-fielding; off-source images (or the dithered ones for standard stars)
were combined together and used for flat-fielding in the $J$ and $H$ bands.
All images in each band
were finally corrected for bad-pixels, registered and averaged together.

Late time NIR images were acquired with NICS at the TNG on December 14, 2000 (JD+893) during the
commissioning of the camera. NICS is based on a $1024 \times 1024$ Hawaii detector 
(Baffa et al.\ 2001) and we
employed a $0.25$ arcsec/pixel scale resulting in a field of view of 
$4.2 \times 4.2$ arcmin$^{2}$. Dithered on-source and off-source frames were taken
in the $J_{s}$ and $K'$ bands with total (on-source) integration times
of $1200$ s and $420$ s, respectively. Data were reduced as described above, but no flat-field
frames were available for $K'$ images.

The photometric measurements on SWIRCAM images were performed by PSF fitting
using the DAOPHOT package  in MIDAS  in order to derive relative photometry between the
supernova and  a set of comparison stars
properly selected in the field
among the ones with the highest S/N ratio (1 to 9 in Fig.~\ref{gal:sn}), of which those
calibrated with the TNG data (see below) are a subset.
For these, variability
could be ruled out on the basis of an accurate image-by-image
check of the fluctuations in their relative magnitudes.
PSF fitting photometry was also needed owing to the
partial convolution of the SN with the sources c1, c2 and c3
of Fig.~\ref{gal:sn}.

The presence of the parent galaxy background has been taken into
account by evaluating its influence on the derived photometry of the
SN. In particular cases, i.\ e.,  when the S/N ratio was low,
PSF fitting photometry was carried out using
the ROMAFOT package in MIDAS, by means of which the background gradient
due to the diffuse emission from the galaxy was
accounted for by adopting a tilted plane.

Both aperture and PSF fitting photometry were performed on the ARNICA 
images for field stars using the package DAOPHOT in IRAF (Stetson 1987). 
However, we obtained more satisfactory PSF fits to \el\
using the package ROMAFOT in MIDAS. The 
group of 3 stars close, in projection, to the supernova, all resolved on the TNG images
(see Fig.~\ref{gal:sn}), is composed of, following the nomenclature used in the figure, 
c2 ($K_{\rm s} = 14.97 \pm 0.03$), $\sim 2$ arcsec east of the SN,
c3 ($K_{\rm s} = 16.75 \pm 0.32$), $\sim 2$ arcsec west of the SN, and c1
($K_{\rm s} \sim 19$).
The observed standard stars were used to calibrate both the SN and
the field stars 2 and 9 (see Fig.~\ref{gal:sn}) which served as references for the photometry on 
SWIRCAM frames. $K_{\rm s}$ magnitudes were converted into
$K$ magnitudes by extrapolating the continuum flux within the spectral
region ($\sim 0.1$ $\mu$m wide) of $K$ not covered by the $K_{\rm s}$ filter,
with corrections amounting to $\leq 0.02$ mag for the standard stars and to
$0.06$ mag for \el\ (the uncorrected sources being fainter).
Possible CO molecular line emission falling out of $K_{\rm s}$ but
within $K$, and sometimes occurring
in the case of Type IIn SNe,  was therefore not detected and could not
account for any fraction of the measured flux.
The photometry of the standard stars for the 5 positions
on the array indicates that residual errors of $\sim 0.05$, $\sim 0.04$ and
$\sim 0.02$ mag in the $J$, $H$ and $K_{\rm s}$ bands, respectively,
are still present owing to flat-field inaccuracies.

Differential photometry on the NICS images was performed by PSF fitting using 
ROMAFOT in MIDAS. By inter-comparing instrumental magnitudes obtained both for the
field (and local standard) stars and for the close-by sources (c1-c3), we checked that
even though not flat-fielded, $K'$ values where consistent at a $0.1-0.2$ mag level with
ARNICA photometry throughout
the whole final frame. We conservatively assumed uncertainties $\sim 0.5$ mag both for
$J_{s}$ and for $K'$ and did not correct for the slight differences with the $J$ and $K$ bands. 
Table~\ref{nir:phot} lists the results of NIR photometry for \el.

%
%
\begin{deluxetable}{ccccccccc}  
\tablecaption{NIR photometry of \protect\el\ obtained with SWIRCAM, ARNICA and 
           NICS. \label{nir:phot}}  
\tablewidth{0pt}
\tablehead{
\colhead{Julian day} & \colhead{Epoch} & \colhead{$J$} & \colhead{$\Delta J$} & \colhead{$H$} 
& \colhead{$\Delta H$} & \colhead{$K$} & \colhead{$\Delta K$}& 
\colhead{Telescope} \\  
 \colhead{(2451000+)}& \colhead{(days)} & \colhead{(mag)} & \colhead{(mag)} & 
\colhead{(mag)} & \colhead{(mag)} & \colhead{(mag)} & \colhead{(mag)}& 
\colhead{ }
} 
\startdata
   $477.31$ & 5.31 &  $13.69$ &  $0.03$ &  $13.30$ &  $0.05$ & $12.70$ &  $0.09$&AZT-24    \\  
   $479.25$ & 7.25 &  $13.62$ &  $0.08$ &  $13.25$ &  $0.08$ & $12.68$ &  $0.08$ &AZT-24   \\  
   $480.28$ & 8.28 &    -      &    -     &  $13.28$ &  $0.09$ &    -     &    -    & AZT-24   \\  
   $481.37$ & 9.37 &  $13.58$ &  $0.08$ &  $13.29$ &  $0.09$ & $12.75$ &  $0.09$ &AZT-24    \\  
   $483.41$ & 11.41 &  $13.55$ &  $0.10$ &  $13.25$ &  $0.06$ &    -     &    -     & AZT-24   \\  
   $484.31$ & 12.31 &    -      &    -     &  $13.08$ &  $0.08$ &    -     &    -     & AZT-24   \\  
   $485.41$ & 13.41 &  $13.55$ &  $0.08$ &  $13.28$ &  $0.07$ &    -     &    -     & AZT-24   \\  
   $486.35$ & 14.35 &  $13.51$ &  $0.06$ &     -     &    -     &    -     &    -     & AZT-24   \\  
   $492.32$ & 20.32 &  $13.46$ &  $0.06$ &  $13.18$ &  $0.15$ &    -     &    -     & AZT-24   \\  
   $493.28$ & 21.28 &  $13.53$ &  $0.08$ &  $13.40$ &  $0.09$ & $12.94$ &  $0.06$  & AZT-24   \\  
   $494.50$ & 22.50 &  $13.43$ &  $0.07$ &     -     &    -     &    -     &    -      &AZT-24    \\  
   $495.20$ & 23.20 &  $13.54$ &  $0.07$ &  $13.36$ &  $0.08$ &    -     &    -      & AZT-24   \\  
   $506.45$ & 34.45 &  $13.73$ &  $0.07$ &  $13.52$ &  $0.07$ & $13.17$ &  $0.05$  & AZT-24  \\  
   $507.23$ & 35.23 &  $13.72$ &  $0.07$ &     -     &    -     &    -     &    -    &  AZT-24  \\  
   $514.25$ & 42.25 &  $13.85$ &  $0.10$ &  $13.77$ &  $0.09$ & $13.30$ &  $0.06$&  AZT-24   \\  
   $521.25$ & 49.25 &  $13.96$ &  $0.05$ &     -     &    -     & $13.49$ &  $0.07$ & AZT-24   \\  
   $531.71$ & 59.71 &    -      &    -     &  $14.24$ &  $0.09$ &    -     &    -    & AZT-24   \\  
   $532.28$ & 60.28 &  $14.24$ &  $0.08$ &     -     &    -     & $13.91$ &  $0.08$ & AZT-24   \\  
   $548.21$ & 76.21 &  $14.88$ &  $0.04$ &     -     &    -     & $14.40$ &  $0.05$&  AZT-24   \\  
   $549.22$ & 77.22 &  $14.92$ &  $0.03$ &     -     &    -     & $14.47$ &  $0.06$ & AZT-24    \\  
   $550.24$ & 78.24 &  $14.99$ &  $0.04$ &  $14.75$ &  $0.15$ & $14.53$ &  $0.04$ & AZT-24    \\  
   $576.24$ & 104.24 &  $17.20$ &  $0.30$ &  $16.60$ & $0.30$ & $16.00$  & $0.20$ & AZT-24 \\
   $585.71$ & 113.71 &  $17.50$ & $0.50$ & $16.60$   & $0.40$ & - & - & AZT-24 \\
   $624.59$ & 152.59 &  $17.70$  &  $0.60$ &  $17.60$ &  $0.50$   & $16.40$   & $0.60$ & AZT-24    \\   
   $691.62$ & 219.62 &  $18.31$  &  $0.40$ &  $17.84$ &  $0.50$  & $16.43$  &  $0.25$  & TNG/ARNICA  \\  
   $893.34$ & 421.34 &  $20.01$  &  $0.50$ &     -    &    -     & $17.40$  &  $0.50$  & TNG/NICS  \\  
\enddata
\end{deluxetable}  
  
\subsection{Spectroscopy} 
 
\begin{figure}  
\epsscale{0.70}
\plotone{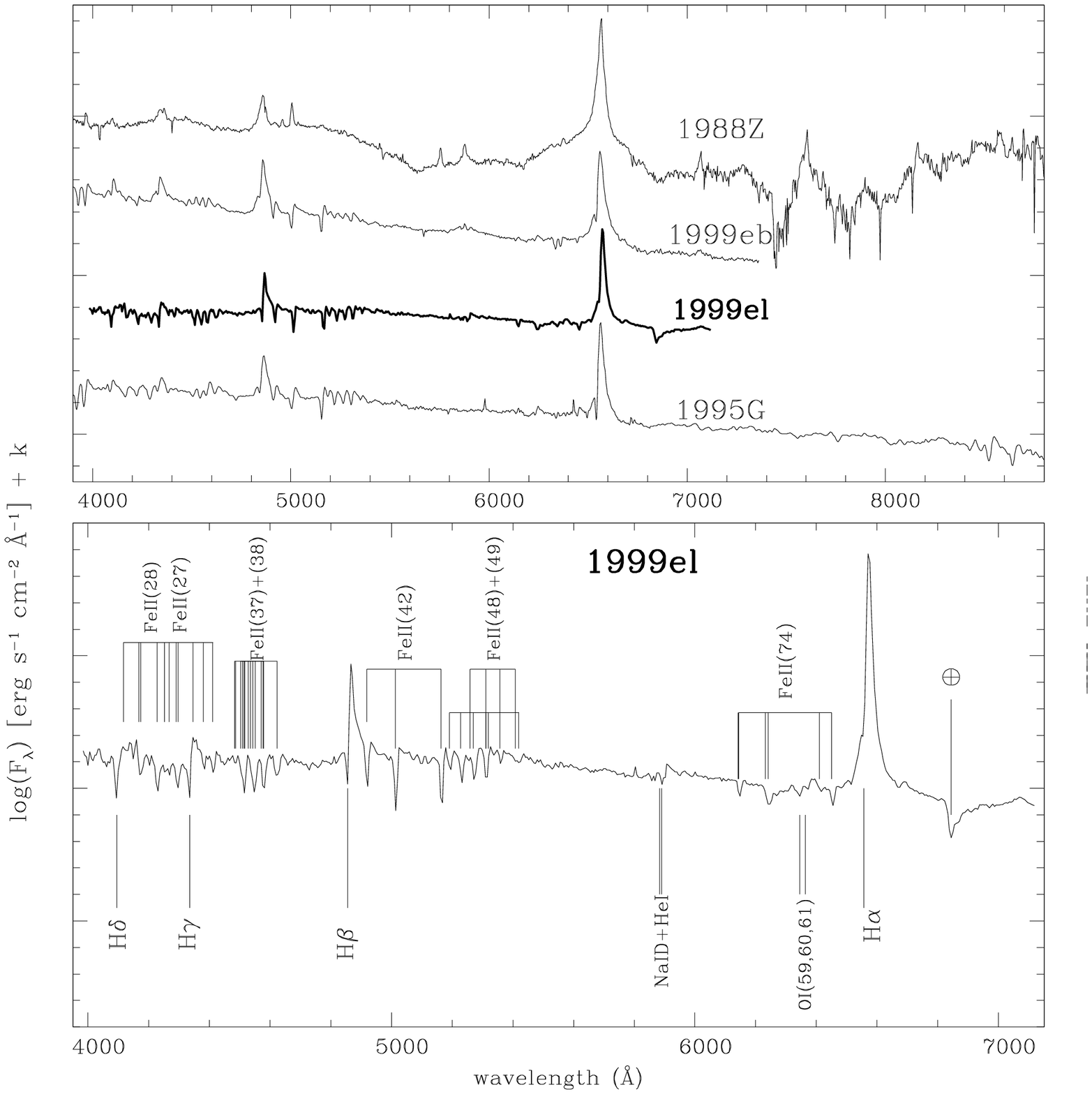}  
\caption{Identification of the lines 
  on the optical spectrum of SN~1999el obtained on November 8, 1999, in Cananea 
  (lower panel). Comparison with other SN spectra is shown  in the upper panel.  
  Vertical bars indicate the position of absorption lines blue-shifted by $\sim 600-900$
  km s$^{-1}$ with respect to the emission lines cores.
\label{spectrum}}  
\end{figure}
 
Two spectra of the SN have been obtained a few hours apart in Asiago 
(November 7, 1999) and Cananea (November 8), i.e. about 18 days after the discovery. 
In Fig.~\ref{spectrum} we show the latter obtained with a 5.5 
\AA/pixel grism, which for a $3''$ slit-width, yields a resolution of 
$\sim 16$ \AA\ in the 4000--7200\AA\ range. The total 4400~s exposure 
was split into 3 integrations to minimize the effect of cosmic rays. 
Wavelength calibrations were obtained using 
comparison spectra of HeArNe lamps, while the instrumental signature 
has been removed using a spectrophotometric standard star. 
 
\section{Results and analysis} 

\subsection{Optical spectrum}
\label{spec:res}

The spectrum (Fig.~\ref{spectrum}) confirms the early finding by Filippenko (1999) 
that the SN shows lines with narrow components. Superimposed on
the continuum there are strong, 
relatively narrow Balmer emission lines of H together with wider weak wings. 
The broad wings could be fitted with Gaussian 
profiles having FWHM of about 2500--3000 \kms, while the narrow ones are 
barely resolved. Also narrow P-Cyg absorption components with minima 
displaced by about 600--900 \kms\ relative to the emission-line cores 
are present, 
indicating a slowly expanding shell of gas above the photosphere.  
Fe II lines have similar structures. Weak NaI D lines are visible. 
 
In Fig.\ref{spectrum} we compare the spectrum of SN~1999el with those of 
other H--dominated SNe at similar epochs showing signs of interaction with the 
CSM. The SN 1999el spectrum is very different from the spectra of SN~1988Z, 
the best studied Type IIn SN (Stathakis \& Sadler 1991; Turatto et 
al. 1993; Chugai \& Danziger 1994; Aretxaga et al. 1999). The line widths of SN~1988Z are much 
broader, implying expansion velocities of about 15000 \kms. 
On the contrary the spectrum 
of SN 1999el closely resembles those of SNe 1999eb and 1995G (Pastorello et
al.\ 2002, in preparation) at comparable age. They have similar continua, strong 
emission lines with similar profiles and narrow absorption components. It is the very 
narrow blue shifted absorption component particularly in the
wing of the broad H$\alpha$ from the expanding envelope that is characteristic of a class of SNe, 
such as SN~1996L (Benetti et al. 1999), SN 1994aj (Benetti et al. 1998) and SN~1984E 
(Dopita et al.\ 1984). This absorption is commonly thought to arise from the CSM in the form 
of a wind.
 
\subsection{Light curves}
\label{lc:d}

%
%
\begin{figure}
\epsscale{0.8}
\plotone{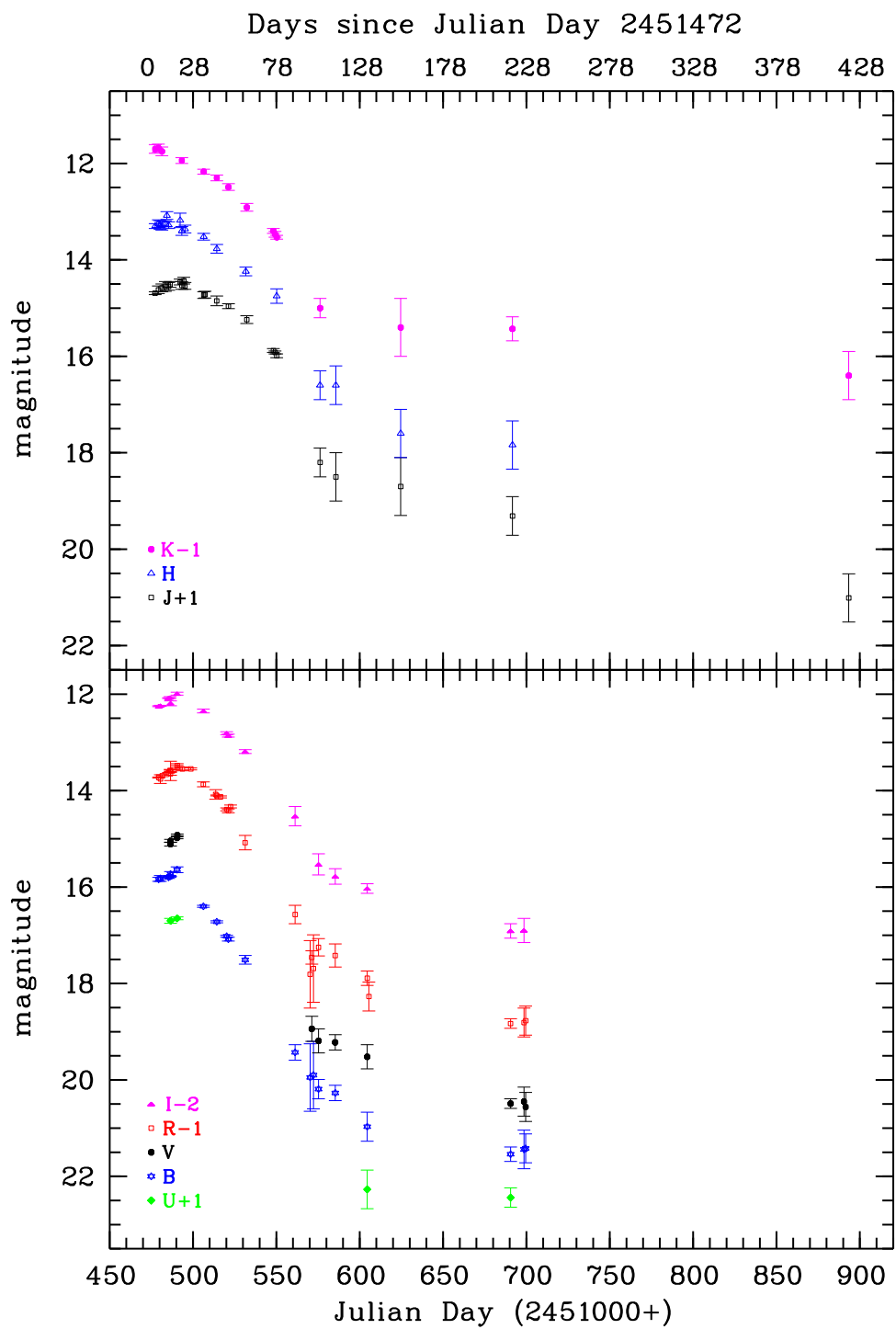}
\caption{Optical and infrared light curves of the SN~1999el. Errors marked
 by the bars are those indicated in Table~\ref{Table1} and \ref{nir:phot}.  
\label{f:fig2}}  
\end{figure}   
%
 
Figure \ref{f:fig2} shows the LCs at  
optical and NIR bands; we draw attention to some particular features.  
Firstly, we see that 
a maximum of emission in the $K$ band has not been observed.  
However, it seems that LC maxima in the other  
NIR and optical bands have been recorded. There  
are indications of small  delays in the time of maxima occurrence 
which increase with a decrease of wavelength. During the  
first 90 days, the decline rate is quite fast reminiscent  
of that of a Type II-L SN rather than that of a typical IIn SN,
as already noted.   
After 90 days the LCs exhibit a much slower decrease and, in  
particular, that of the $K$ band flattens most.
Since the $K$ point at JD+690 has been derived from a  
$K_{\rm s}$ measurement and any possible line contribution  
from CO emission is therefore excluded, the late behavior of
the LC in $K$ is not due to CO line emission. 
Note that this point has been 
derived from TNG observations, so the SN and its nearby sources are well resolved.   
The NIR LCs appear to remain flat up to JD+893 ($\sim 420$ days from the shock outbreak),
even if in the $J$ band the decrease seems slightly faster than in the $K$ band.  
 
%
%
\begin{figure}  
\epsscale{0.5}
\plotone{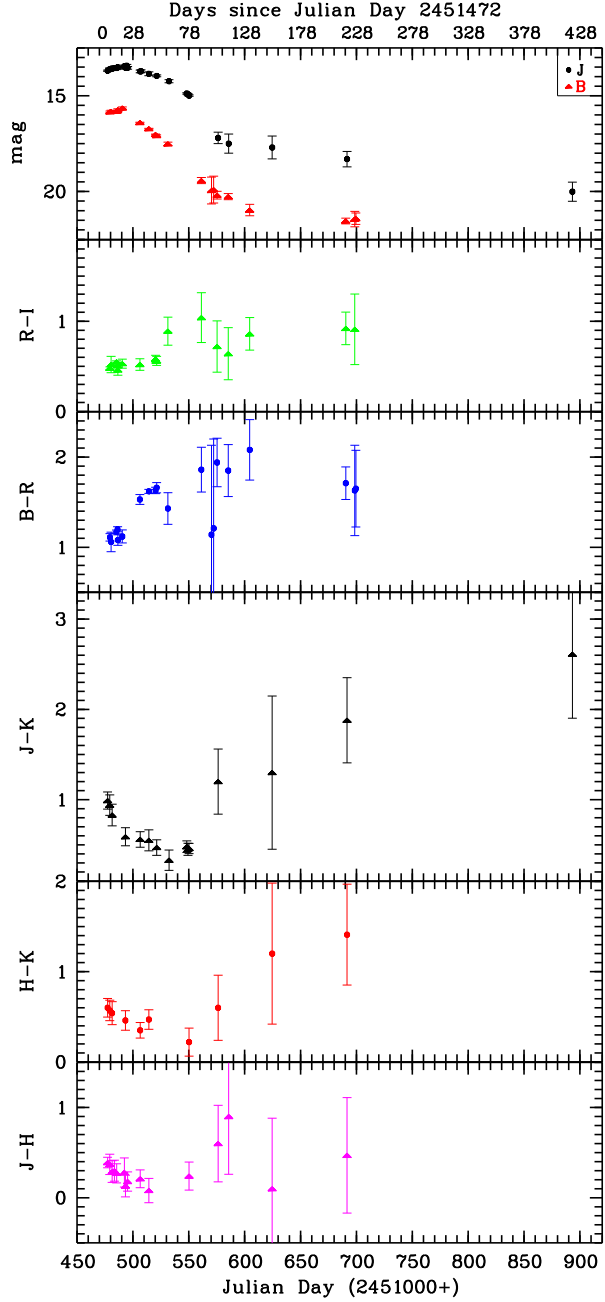}
\caption{Optical and infrared colors of the SN~1999el. For the sake of
 comparison, in the upper box an optical ($B$) and an infrared ($J$) LC
 are displayed. 
\label{f:fig3}}  
\end{figure}   
  
Figure \ref{f:fig3} shows the time evolution of a series of color indices.
We note the blueing in  
$J-H$ and $J-K$ during the first $\sim 45$ days (including  
$\sim 10$ days before the occurrence of maximum in the $J$ band).  
Afterwards, starting approximately around JD+550, 
a reddening in $J-K$ and $J-H$ colors begins to be evident. The increase 
in errors which is significant after JD+550 (affecting all colors) 
is due to the SN magnitude becoming comparable
with that of the brightest of the nearby field stars,  
the two sources being difficult to deconvolve on the images
from the smaller telescopes,
even though we already noted that the $K$ point obtained at late times from TNG 
images (JD+690) is well established.  This reddening 
effect is also evident as a flattening of the $K$ LC 
whereas $J$ and $H$ 
are still slightly steeper in their decline (see Fig.~\ref{f:fig2}) 
and can be attributed to an infrared excess arising at later phases. 
\el\ exhibits a more standard behavior 
in $B-R$, which increases after maximum and then flattens after $\sim 70$
days. This may be understood in terms of a cooling 
photosphere. In fact, the same trend seems to  
characterize all optical colors, with a change of $\sim 0.5$ mag  
in a period spanning $\sim 200$ days. 
 
\begin{figure}
\epsscale{0.8}
\plotone{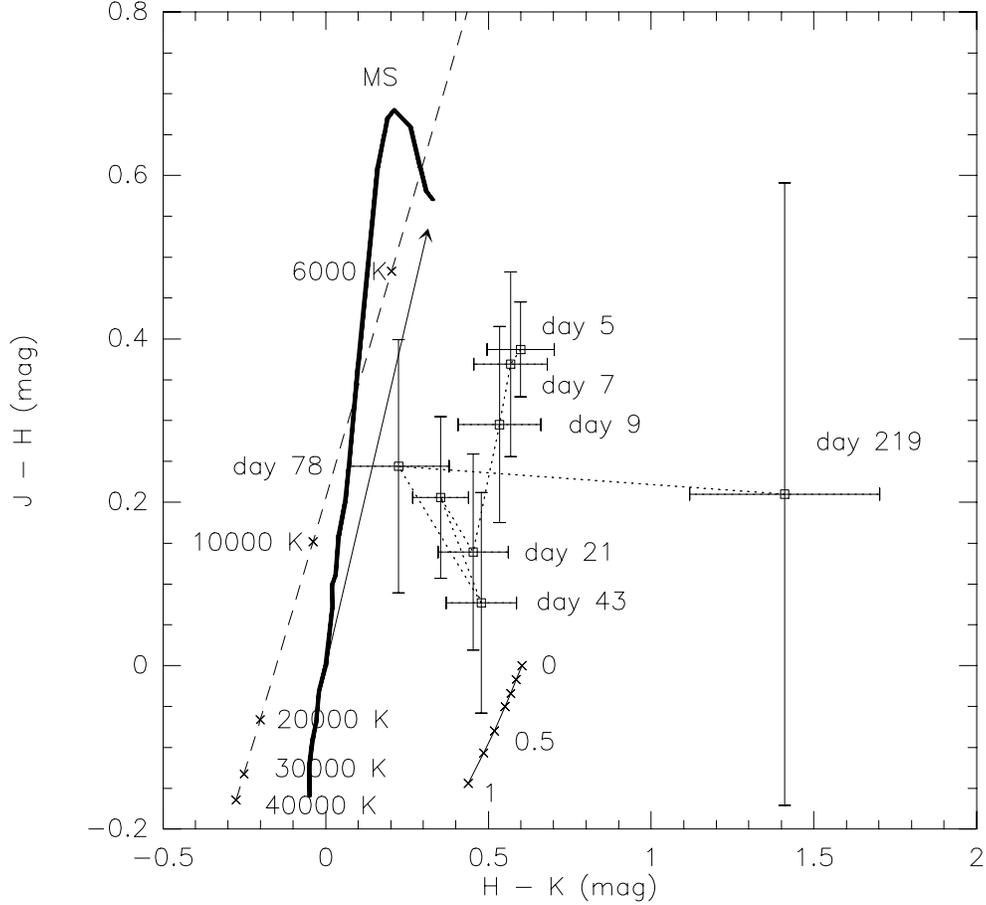}  
\caption{Color-color diagram ($H-K$ vs.\ $J-H$) showing 
   the time evolution of the NIR emission of SN~1999el. 
   The open squares (with error bars) mark the measured colors,
   which evolve along the dotted line from the 5th day from
   the discovery to the 219th day (the epoch is indicated
   near each data point). 
   The solid black line reproduces the main sequence locus
   from O6--8 to M8 stars (according to the colors given
   by Koornneef 1983), 
   whereas the arrow from $(0,0)$ corresponds to a reddening
   A$_{V}$ $ = 5$ mag according to the standard law of
   Cardelli et al. (1989).
   The dashed line indicates blackbody colors with crosses  
   at T=40000, 30000, 20000, 10000 and 6000 K.
   The effect of including FeII line emission at $1.26$ and 
   $1.64$ $\mu$m superimposed on a NIR continuum 
   (assumed from the $JHK$ values on the 5th day)
   is shown by the solid line below the point of the 43rd day. 
   The shifts are labeled in units (0,0.5,1) of the ratio of
   line integrated emission at $1.26$ $\mu$m to 
   the overall $K$-band integrated continuum flux. For the sake of
   clarity, the line has been moved downward by $0.395$ mag
   (i. e., the mark labeled as 0 actually falls over the point
   of day 5). 
\label{f:cc}}  
\end{figure}

The behavior of the NIR emission can be followed on a color-color 
diagram ($H-K$ vs.\ $J-H$; see Fig.~\ref{f:cc}). Comparing the 
SN colors with the main sequence locus (solid line), the blackbody 
locus (dashed line) and the reddening law (arrow), 
a NIR excess is evident even at early epochs 
and this cannot be simply explained as  reddened 
blackbody or stellar photospheric emission. After day 5, the colors 
evolve roughly along a track which appears almost parallel to the 
reddening law, evolving as if A$_{V}$ decreases with time. 
However, on day 219, a strong $K$ excess is present.
The most likely cause for the NIR excess is continuum emission from hot
dust pre-existing within the CSM of \el. Following Fassia et al. (2000),
we can test this scenario in the following way. Given the (reddened) flux in the 
$K$ band 
for day 5, $3.39 \times 10^{-12}$ erg cm$^{-2}$ s$^{-1}$ $\mu$m$^{-1}$,
and the distance assumed in Sect.~\ref{bbfs}, we can determine the radius of
a blackbody surface emitting the same luminosity in $K$ at 1500 K, 
roughly the dust sublimation temperature. 
It can be shown that reddening does not change the result by more than a
10--20\%. This radius, $1.5 \times 10^{16}$ cm, is
comparable to the light travel distance from the explosion time to 
day 5, $\sim 1.3 \times 10^{16}$ cm. 
Thus the two independently determined distances are consistent with the presence of dust
at a radius where it has not sublimated and radiates at IR wavelengths as a result of
heating by UV photons from the original outburst. 
 
%
%

\subsection{Reddening} 
 
In order to give an estimate of the absolute luminosity 
of \el\ it is necessary to determine its reddening. 
As can be seen in Fig.~\ref{spectrum}, 
the interstellar lines of NaI cannot be resolved and therefore add little useful
information. Hence, we can only 
assume the galactic value as a lower limit. From the maps of Schlegel, Finkbeiner, \& 
Davis (1998) we  
obtain A$_{B}$ $ = 1.57$ mag. This map appears to systematically
overestimate  the extinction where $E(B-V) \geq 0.15$ mag, as
pointed out by  Arce \& Goodman (1999). 
Since this occurs towards \el, as well, we expect a lower galactic contribution
to the extinction.  We dereddened optical and NIR magnitudes 
using the interstellar extinction law given by Cardelli, Clayton, \& Mathis (1989) and assuming 
an A$_{V}$ ranging between $1.18$ (corresponding to A$_{B}$ $ = 1.57$ mag) 
and $2.5$ mag. These were then converted to fluxes according to Bessel (1979) 
and, in the NIR bands, the calibration given for the UKIRT standard. In Table~\ref{all:max}
both apparent and absolute (dereddened) magnitudes at maxima are listed. 

%
%
\begin{deluxetable}{cccc}
\tablecaption{Apparent and absolute (dereddened) magnitudes for all observed
         photometric bands. \label{all:max}}
\tablewidth{0pt}
\tablehead{
\colhead{ } & \colhead{ } & \colhead{A$_{V}$ $=1.18$} & \colhead{A$_{V}$ $=2.50$} \\
\colhead{Band} & \colhead{Rel.\ mag} & \colhead{Abs.\ mag} & \colhead{Abs.\ mag}
}
\startdata
$U$ & 15.7 & $-18.3$ & $-20.3$ \\
$B$ & 15.6 & $-18.1$ & $-19.8$ \\
$V$ & 15.0 & $-18.3$ & $-19.6$ \\
$R$ & 14.5 & $-18.5$ & $-19.5$ \\
$I$ & 14.0 & $-18.7$ & $-19.3$ \\
$J$ & 13.4 & $-19.0$ & $-19.4$ \\
$H$ & 13.1 & $-19.2$ & $-19.5$ \\
$K$ & 12.7 & $-19.5$ & $-19.7$ \\
\enddata
\end{deluxetable}

\subsection{Blackbody fits}
\label{bbfs}
\begin{figure} 
\epsscale{0.5}
\plotone{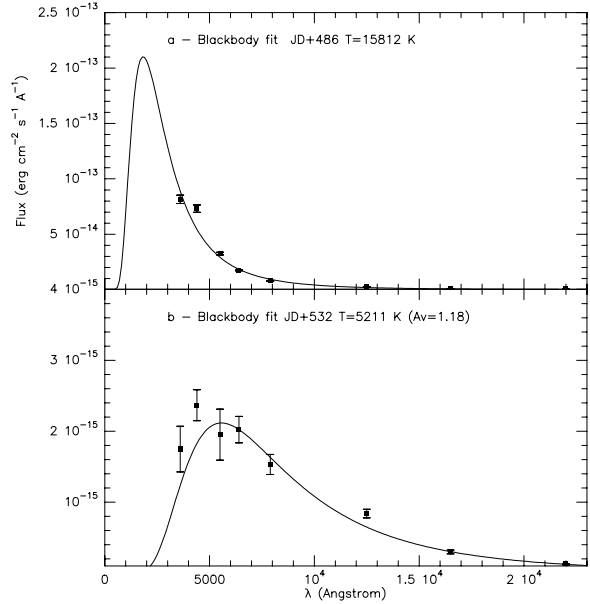} 
\caption{Blackbody fits to the dereddened fluxes for ({\rm a}) JD+486 
(assuming A$_{V}$ $ = 2.50$ mag) and ({\rm b}) JD+532 
(assuming A$_{V}$ $ = 1.18$ mag).  
\label{f:fits}} 
\end{figure}

We fitted a blackbody law to the 
dereddened fluxes in order to get estimates of the photospheric temperature. Since 
not all bands are covered on the same days, we constructed sets of 
$UBVRIJHK$ fluxes complementing the available values with ones 
obtained through interpolation or using the results of 
spline fits to all data in a given wavelength. Then, we adopted a 
one-parameter procedure for $\chi^{2}$ minimisation normalizing, 
for each selected day, all fluxes to that in the $R$-band in 
order to remove the dependence on radius and distance. $R$ was chosen because 
it is the better sampled band  (see Table~\ref{Table1}) and is
likely to be much less affected by dust emission than the NIR bands.      
Two of the fits are shown in Fig.~\ref{f:fits} as examples of earlier and later 
epoch emission. 
 
%
%
 
\begin{figure}  
\epsscale{0.50}
\plotone{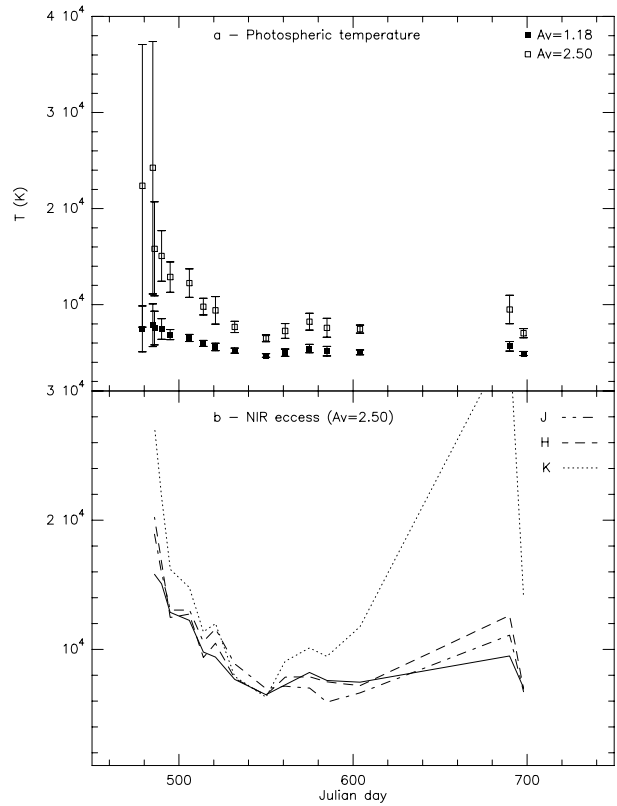}  
\caption{{\bf a} Photospheric temperatures for 
  different assumptions on the extinction and {\bf b} 
  blackbody temperatures obtained from the NIR fluxes 
  (for A$_{V}$ $=2.5$ mag) 
  scaled using a diluition factor equal to the ratio of 
  observed to blackbody flux in the $R$ band, compared 
  to the photospheric temperature (solid black line).  
\label{f:temp}}  
\end{figure}

The resulting photospheric temperatures are listed in Table~\ref{temp} and shown in 
Fig.~\ref{f:temp}a for A$_{V}$ $=1.18$  and
$2.5$ mag.
The temperatures for A$_{V}$ $=2.50$ mag are similar to those derived in the 
case of SN~1998S (Fassia et al.\ 2000); an extinction as large as 
A$_{V}$ $=3.00$ mag drives the photospheric temperature
at early phases towards values which 
appear too high.
Hence, A$_{V}$ cannot much exceed $\sim 2.5$ mag and 
A$_{V}$ $=1.18$ mag can be safely assumed as a lower limit.  
 
%
%
\begin{deluxetable}{cccc}  
\tablecaption{Photospheric temperatures from blackbody curve 
         fits for different extinctions. \label{temp}}  
\tablewidth{0pt}
\tablehead{
\colhead{Julian Day} & \colhead{Epoch} & \colhead{A$_{\rm V}$ $=1.18$} & \colhead{A$_{V}$ $=2.50$} \\  
\colhead{ } & \colhead{(days)}  & \colhead{$T$(K)} & \colhead{$T$(K)} 
}
\startdata
479 & 7 & $7468  \pm 2390$ & $22370 \pm 14717$ \\               
485 & 13 & $7852  \pm 2217$ & $24254 \pm 13135$ \\               
486 & 14 & $7561  \pm 1758$ & $15812 \pm 4901$ \\               
490 & 18 & $7462  \pm 1065$ & $15066 \pm 2629$ \\                
495 & 23 & $6877  \pm 534$ & $12875 \pm 1579$ \\               
506 & 34 & $6533  \pm 369$ & $12245 \pm 1488$ \\               
514 & 42 & $5957  \pm 310$ & $9791 \pm 871$ \\               
521 & 49 & $5591  \pm 389$ & $9405 \pm 1429$ \\               
532 & 60 & $5211  \pm 262$ & $7683 \pm 587$ \\               
550 & 78 & $4699  \pm 182$ & $6488 \pm 350$ \\                   
561 & 89 & $4997  \pm 377$ & $7262 \pm 774$ \\                 
575 & 103 & $5419  \pm 441$ & $8220 \pm 874$ \\               
585 & 113 & $5145  \pm 498$ & $7586 \pm 987$ \\               
604 & 132 & $5029  \pm 254$ & $7457 \pm 419$ \\               
690 & 218 & $5660  \pm 489$ & $9487 \pm 1480$ \\                
698 & 226 & $4910  \pm 205$ & $7026 \pm 474$ \\                
\enddata
\end{deluxetable} 
 
We performed several tests in order to evaluate the quality of our 
fits. Most important, we repeated our procedure normalizing all fluxes to that 
in the $V$ band. The differences between these newly obtained values 
and those listed in Table~\ref{temp} increase with A$_{V}$, 
ranging from $< 6$\% for A$_{V}$ $ = 1.18$ mag, to $< 20$\%  
for A$_{V}$ $ = 2.50$ mag.
The most sensitive variations are observed on JD+495  
(temperatures from $V$-normalized data hotter by 
23\% for A$_{V}$ $ = 2.50$ mag 
than those from $R$-normalized data) and on JD+506 (with a similar trend). 
Here, two kinds of problems exist. First, when $\lambda T >> hc/k$ 
a $\lambda_{0}$-normalized blackbody flux tends to become equal to 
$(\lambda_{0}/\lambda)^{4}$ and no longer depends on temperature. 
This occurs for $T >> 30000$ K in the $V$-band and $T >> 40000$ K in the 
$U$-band. Then, our fits are not reliable for very high temperatures.
Second, differences in the results obtained by varying  the  
normalization band may be related to the general increase in 
temperature found when excluding NIR fluxes from the fits (amounting, 
e.\ g., to up to 60\% for A$_{V}$ $ = 2.50$ mag), although the trend shown 
in Fig.~\ref{f:temp}a remains unchanged. This reflects physical differences in
the regions which emit the bulk of radiation at different wavelengths
(particularly, in the optical and in the near-infrared).
As an example, for SN~1998S 
Anupama, Sivarani, \& Pandey (2001) when fitting blackbody curves to optical spectra
find photospheric temperatures up to 30 \% hotter  
than those obtained by Fassia et al.\ (2000) from photometric 
data including NIR fluxes.
We also repeated our fitting procedure excluding
fluxes in the $U$ band, where we have very few measurements.
This yields somewhat higher temperatures, but  
only at early epochs (20 \% on JD+486 and 60\% on JD+490 
for A$_{V}$ $ = 2.50$ mag); thereafter, differences amount to only a few percent.   
A major source of uncertainty is line emission in the photometric bands.
Particularly at later stages ($> 100$ days from LCs maxima), lines affecting
the photometry in the $I$ and $R$ bands may develop (see Danziger et al. 1988)
causing temperature underestimates.

%
%
  
The errors listed in Table~\ref{temp} (calculated as the square root 
of the quadratic sum of the temperature residuals 
in the various bands) are largely dominated 
by those involving the NIR fluxes. We have already discussed the indication of 
a NIR excess from $JHK$ colors. This is further demonstrated by plotting the 
blackbody temperatures obtained scaling the dereddened fluxes 
in the NIR to the ratio of observed to blackbody fluxes in the 
$R$-band (which is equal to the square of the ratio of  
photospheric radius to 
distance). As shown in Fig.~\ref{f:temp}b, higher NIR temperatures 
are derived at both early and late epochs. This may be 
reconciled if the NIR fluxes are not photospheric in origin but come from 
larger (cooler) regions starting soon after the explosion and continuing.   
Interestingly, an excess in the $J$ band is exhibited until  
JD+560 and is quite evident during the period (JD+)520--550 
(see also Fig.~\ref{f:fig2}); it corresponds to the 
noted blueing at early days. The $K$-band excess at late 
epochs is very strong, so much so that the flux has been left out when evaluating  
the error for JD+604, JD+690 and JD+698. 
Fluctuations in the derived photospheric  
temperatures are appreciable in Fig.~\ref{f:temp}a and may be 
related to the undersampling of LCs in some bands (e.\ g., $U$ 
and $V$). Unaccounted for line emission may also contribute. 
The apparent warming during the period (JD+)550--600 
coincides with the fading of the SN towards brightness values similar  
to those of the brightest and closest star and it may indicate 
that the two sources could not be accurately deconvolved at this epoch.
 
In conclusion, the photospheric temperatures derived from blackbody fits  
between (JD+)486--690 are mostly 
affected by uncertainties related to the unknown extinction,
line emission within the photometric bands 
and the degree of NIR excess. We believe that assuming  
A$_{V}$ $ = 1.18$ mag and A$_{V}$ $ = 2.50$ mag should result 
in lower and upper limits whose errors are  
otherwise not larger than those 
intrinsic to other similar studies (such as Fassia et al.\ 2000). 
Most significantly, the temporal evolution appears 
well defined.  
 
\subsection{Physical parameters of the SN photosphere}

\begin{figure}  
\epsscale{0.8}
\plotone{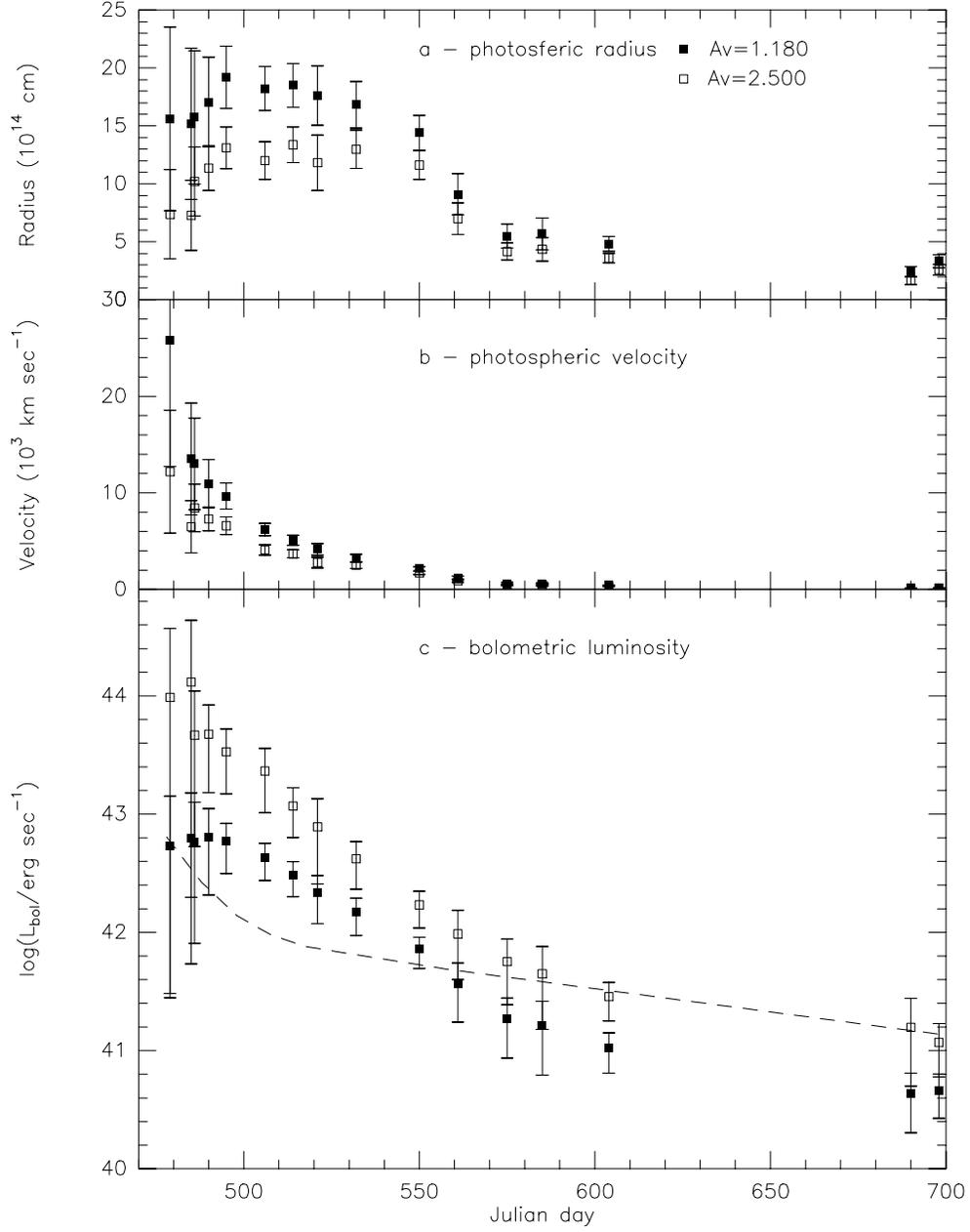}  
\caption{{\bf a} Photospheric radius,  
  {\bf b} photospheric velocity and {\bf c} bolometric 
  luminosity for different assumptions on the extinction.  
  The dashed curve shows the radioactive decay luminosity of 
  $0.07$ M$_{\odot}$ of $^{56}$Ni. 
\label{f:rvl}}  
\end{figure}

The photospheric radius was determined from the ratio of blackbody 
to observed fluxes in the $R$-band. We slightly revised the 
distance to the host galaxy, assuming $26.3 \pm 4.8$ Mpc, from the 
distance modulus ($32.1 \pm 0.4$ mag) we derived from our own 
observations of SN~2000E. Once the photospheric radius and temperature 
are known, the absolute bolometric luminosity can be estimated 
through the Stefan-Boltzmann law. A photospheric velocity was 
also derived as the ratio of radius to elapsed time from epoch 
0. These data are listed in Table~\ref{lboltab} in the cases 
A$_{V}$ $=1.18$ mag and A$_{V}$ $=2.5$ mag and shown in Fig~\ref{f:rvl}.   
Quoted errors include the uncertainties in photospheric temperatures 
and measured fluxes, but not on the distance, since this would affect 
all data similarly.
Note that increasing the photospheric temperature, 
e.\ g., by assuming larger extinctions, produces smaller radii and 
velocities; this confirms that A$_{V}$ cannot much exceed the galactic 
value and rules out the occurrence of very high photospheric temperatures. 
 
\begin{deluxetable}{cccccccc}
\tabletypesize{\scriptsize}
\tablecaption{Photospheric radius, velocity and absolute luminosity
         for different extinctions. \label{lboltab}} 
\tablewidth{0pt}
\tablehead{
           & \multicolumn{3}{c}{A$_{V}$ $=1.18$} & \multicolumn{3}{c}{A$_{V}$ $=2.50$} \\  
\colhead{Julian day} &  \colhead{Epoch} & \colhead{Radius} & \colhead{Velocity} & 
\colhead{Bol.\ luminosity} &  \colhead{Radius} & 
\colhead{Velocity} & \colhead{Bol.\ luminosity} \\ 
 \colhead{ } & \colhead{(days)} & \colhead{$10^{15}$ cm} & \colhead{$10^{3}$ km s$^{-1}$} & 
\colhead{$10^{43}$ erg s$^{-1}$} & 
\colhead{$10^{15}$ cm} & \colhead{$10^{3}$ km s$^{-1}$} & 
\colhead{$10^{43}$ erg s$^{-1}$}
}
\startdata
479 & 7 & $1.6 \pm 0.8$ & $25.81 \pm 13.09$ & $0.54^{+0.88}_{-0.51}$ & $0.74 \pm 0.39$ & $12.20 \pm 6.37$ & $9.71^{+27.5}_{-9.68}$ \\  
485 & 13 & $1.5 \pm 0.7$ & $13.52 \pm 5.80$ & $0.62^{+0.89}_{-0.57}$ & $0.73 \pm 0.30$ & $6.49 \pm 2.70$ & $13.1^{+30.4}_{-12.9}$ \\  
486 & 14 & $1.6 \pm 0.6$ & $13.01 \pm 4.75$ & $0.58^{+0.68}_{-0.50}$ & $1.0 \pm 0.3$ & $8.45 \pm 2.46$ & $4.7^{+6.4}_{-4.1}$ \\ 
490 & 18 & $1.7 \pm 0.4$ & $10.97 \pm 2.49$ & $0.64^{+0.47}_{-0.44}$ & $1.1 \pm 0.2$ & $7.30 \pm 1.23$ & $4.7^{+3.7}_{-3.2}$ \\ 
495 & 23 & $1.9 \pm 0.3$ & $9.67 \pm 1.35$ & $0.59^{+0.25}_{-0.27}$ & $1.3 \pm 0.2$ & $6.60 \pm 0.91$ & $3.4^{+1.9}_{-1.9}$ \\ 
506 & 34 & $1.8 \pm 0.2$ & $6.21 \pm 0.65$ & $0.43^{+0.13}_{-0.15}$ & $1.2 \pm 0.2$ & $4.09 \pm 0.55$ & $2.3^{+1.3}_{-1.3}$ \\ 
514 & 42 & $1.9 \pm 0.2$ & $5.10 \pm 0.52$ & $0.31^{+0.09}_{-0.11}$ & $1.3 \pm 0.2$ & $3.69 \pm 0.42$ & $1.2^{+0.5}_{-0.5}$ \\ 
521 & 49 & $1.8 \pm 0.3$ & $4.16 \pm 0.60$ & $0.22^{+0.09}_{-0.10}$ & $1.2 \pm 0.2$ & $2.79 \pm 0.56$ & $0.78^{+0.57}_{-0.52}$ \\ 
532 & 60 & $1.7 \pm 0.2$ & $3.25 \pm 0.39$ & $0.15^{+0.05}_{-0.05}$ & $1.3 \pm 0.2$ & $2.51 \pm 0.32$ & $0.42^{+0.17}_{-0.19}$ \\ 
550 & 78 & $1.4 \pm 0.2$ & $2.14 \pm 0.22$ & $0.072^{+0.019}_{-0.023}$ & $1.2 \pm 0.1$ & $1.73 \pm 0.18$ & $0.17^{+0.05}_{-0.06}$ \\ 
561 & 89 & $0.9 \pm 0.2$ & $1.18 \pm 0.23$ & $0.037^{+0.018}_{-0.019}$ & $0.70 \pm 0.13$ & $0.91\pm 0.18$ & $0.10^{+0.06}_{-0.06}$ \\ 
575 & 103 & $0.5 \pm 0.1$ & $0.62 \pm 0.12$ & $0.019^{+0.009}_{-0.010}$ & $0.42 \pm 0.07$ & $0.47 \pm 0.08$ & $0.057^{+0.031}_{-0.032}$ \\ 
585 & 113 & $0.6 \pm 0.1$ & $0.58 \pm 0.14$ & $0.016^{+0.010}_{-0.010}$ & $0.44 \pm 0.10$ & $0.45\pm 0.10$ & $0.045^{+0.031}_{-0.030}$ \\ 
604 & 132 & $0.48 \pm 0.06$ & $0.42 \pm 0.06$ & $0.011^{+0.004}_{-0.004}$ & $0.36 \pm 0.04$ & $0.32 \pm 0.04$ & $0.029^{+0.009}_{-0.011}$ \\ 
690 & 218 & $0.24 \pm 0.04$ & $0.13 \pm 0.02$ & $0.004^{+0.002}_{-0.002}$ & $0.17 \pm 0.03$ & $0.09 \pm 0.02$ & $0.016^{+0.012}_{-0.011}$ \\ 
698 & 226 & $0.33 \pm 0.06$ & $0.17 \pm 0.03$ & $0.005^{+0.002}_{-0.002}$ & $0.26 \pm 0.05$ & $0.13 \pm 0.02$ & $0.012^{+0.05}_{-0.06}$ \\ 
\enddata
\end{deluxetable} 
%

Qualitatively, the photospheric radius, velocity and the bolometric luminosity 
exhibit the same behavior revealed by Fassia et al.\ (2000)  for
SN~1998S. The photospheric radius increases during the first 30 days, 
stays roughly constant until $\sim 80$ days from epoch 0 and then 
declines. The exact amount of the variations depends on the assumed extinction. 
The decrease in the photospheric velocity during the period 30--80 days 
is probably caused by the recession of the hydrogen recombination 
front through the expanding ejecta. However, an in-depth comparison is prevented  
by the uncertainty in the extinction of \el.  
Again, similarly to what was found for SN~1998S by Fassia et al.\ (2000), 
at around $100$ days after outburst the light curve (see Fig.~\ref{f:rvl}c) 
slows to the point of approximating the radioactive decay of $^{56}$Co.    
As shown in Fig.~\ref{f:rvl}c, $0.07$ M$_{\odot}$ of $^{56}$Ni 
(based on the relation given by Branch (1992) or direct comparison with SN~1987A) 
could power the luminosity at later epochs for $A _{\rm V}=2.5$ mag and can 
be assumed as an upper limit to the mass of $^{56}$Ni produced by the explosion.

%
%

\section{Discussion}
\label{disc}
 
The early evolution (within $\sim 100$ days) of the light curves of
\el\ indicates that this
is one of the most rapidly declining Type IIn SN among
those studied so far. It is useful to compare its features with
those of SN~1998S, the other recent and extensively observed Type IIn SN (Fassia et
al.\ 2000).  In the $B$ band, \el\ exhibits a luminosity decrease even faster than
that of SN~1998S and is more similar to a
Type II linear SN than to previously studied
Type  IIn SNe, such as SN~1998Z or SN~1997ab.
The light curves at various wavelengths of \el, with their steep decline,
are closer to those of SNe 1994aj (Benetti et al.\ 1999) and 1996L
(Benetti et al.\ 1998) than to that of SN~1995G (Pastorello et al.\ 2002),
whose spectrum is on the contrary very similar to that of \el,
as noted in Sect.~\ref{spec:res}.
Therefore, the indication of interaction with the CSM
arising from the spectrum is not supported by the rapidly declining light curve.
Consequently, type IIn SNe seem to embrace a wide range of characteristics.
Therefore the differences  between \el\ and SN~1998S are
worth noting.
 
It appears that the SN~1998S LCs attain a maximum
in the NIR some days after those in the optical bands (however, the
precise times of occurrence for the optical maxima are uncertain since measurements
appear to have begun when the optical light curve had
already started declining). Conversely, in the case of \el\
the maximum emission in the NIR $H$ and $K$ bands occurs before that
in the optical bands, whose maxima almost coincide with
that in the NIR $J$ band. The exact time of maxima
at $H$ and $K$ bands is uncertain
but they seem to have occurred some time before the beginning of
our observations.
The differences may be appreciated by looking at
the color curves shown in Fig.~\ref{f:fig3}. As previously noted,
the evolution of $J-H$ and $J-K$ for \el\ exhibits an
early appearance of a NIR excess (around day 5) and a NIR
blueing effect in the first $\sim 45$ days. This seems observationally
well established, since here the NIR LCs sample the SN when it was much
brighter than the nearby sources c2 and c3 (see Fig.~\ref{gal:sn}),
and thus are easily measurable even though sometimes
only partially resolved on SWIRCAM images.
Although maxima at NIR wavelengths do appear to have been missed
for SN~1998S, nevertheless there is no evidence for an early blueing
effect in this case. Furthermore, a NIR excess from SN~1998S is not apparent at
early epochs.
 
One possible origin of the quoted differences,
in particular of the early NIR excess
may be due to an IR echo from pre-existing dust in the vicinity of \el\
not evaporated at the moment of the shock outburst.
This would suggest that dust clumps near \el\ are distributed 
differently than those near SN~1998S which may be devoid of nearby dust 
entirely. For SN~1998S this is supported by 
Leonard et al. (2000) whose spectroscopic observations 
suggest that the progenitor of SN~1998S
underwent a major mass-loss episode that ended some 60 years before the
explosion, followed by a weaker scarcely detectable mass-loss activity
in the last seven years before the explosion. If
a mass-loss wind associated with the
\el\ progenitor had not ceased at the time of
explosion, the density of the CSM around it would be much larger than
in the case of SN~1998S.
 
In the above scenario,
the interaction of the UV flash with the CSM around
\el\ would have led to heating, destruction and sublimation of dust. 
The extent of the region where grain sublimation
and destruction can occur may be estimated as (see, e.\ g., Waxman \& Draine 2000)
\begin{equation}
\label{e:rad}
R_{0} = 3.7 \times 10^{19}(Q_{\rm abs} L_{49} a_{01}^{-1})^{\frac{1}{2}} {\rm cm}
\end{equation}
where $Q_{\rm abs}$ is the grain absorption efficiency at UV wavelengths,
$L_{49}$ the maximum luminosity in the range 1--7.5 eV
in units of $10^{49}$ erg s$^{-1}$ (optical-UV flash) and
$a_{01}$ the mean size of grains in units of $0.1$ $\mu$m.
Assuming $Q_{\rm abs} \sim 1$ and $a_{01} \sim 1$, $R_{0}
\sim 10^{17}$ cm is obtained for a maximum
luminosity of $10^{44}$ erg s$^{-1}$ (the greatest
value indicated in Table~\ref{lboltab}),
a distance which is two orders of magnitude larger
than the largest photospheric radius attained by the supernova
(see Table~\ref{lboltab}), but also somewhat greater than the estimate
based on the occurrence of the NIR excess very early 
(see Sect.~\ref{lc:d}).
Since the grain distruction and sublimation due to the UV flash
imply also the unlocking of Fe ions (if they exist in the dust), 
we have been tempted to
ascribe the early 45 days blueing to [FeII] line emission. 
The [FeII] lines which may affect $J$ and $H$ bands result from the
transitions involving metastable levels
$a^{6}$D$_{9/2}$--$a^{4}$D$_{7/2}$  ($\lambda = 1.26$ $\mu$m, A$=5.6
\times 10^{-3}$ s$^{-1}$)
and $a^{4}$F$_{9/2}$--$a^{6}$D$_{7/2}$  ($\lambda = 1.64$ $\mu$m, A$=
1.9 \times 10^{-3}$ s$^{-1}$), collisionally excited by
free electrons (see Mouri, Kawara, \& Taniguchi 2000, and references therein).
Very large line integrated fluxes (up to 0.5--1 times the continuum flux 
integrated on the $K$ band; see Fig.~\ref{f:cc}) would be required  to 
produce the observed changes in colors,
indicating that [FeII] lines alone cannot account for it. In addition, since 
relative luminosity of the main [FeII] lines in the J and H windows is 
constant, the contribution of this mechanism to the color evolution is 
definitively ruled out.

If the early IR excess of \el\ is due to the presence of a CSM generated 
by a continuous wind of the progenitor star, it 
might be reasonable to expect its signature in the optical light curves. 
This seems not to be the case. 
Roscherr and Schaefer (2000) analysed theoretically the 
modifications of optical LCs supernovae produced by echo processes.
They assume spherical symmetry for the
dust and find that critical parameters are $r_{min}$,
$r_{max}$ and $\tau$, where $r_{min}$ and $r_{max}$ are the inner and
outer radii of the dust shell and $\tau$ the optical depth
in the $R$ band for a purely radial photon trajectory,
a quantity directly related with the total mass of dust.
Roscherr and Schaefer (2000) find that,
in the case of thick shells, $\tau$ is the more sensitive
parameter.
Changes in $\tau$ effect changes in the luminosity of light maxima
in various bands whose amounts depend on the wavelength range.
Even the time at which the light maximum is attained
may be shifted (by up to several days;
Roscherr, private communication) according to its wavelength. 
Unfortunately, these computations do not address the NIR bands.
The decline rate is also strongly affected by changes in $\tau$
at levels once again dependent on wavelength. In this respect
a relevant parameter 
is the $\beta_{100,B}$ index which corresponds to the luminosity
drop in 100 days in the $B$ band. Therefore the $\beta_{100,B}$ index
can provide information on the presence of a CSM around the SN. 
Observational data provide a value of $\beta_{100,B}$ greater than
4 mag for both \el\ and 1998S. A comparison with the theoretical results
quoted above suggests the surprising result 
that the expected amount of dust
surrounding both \el\ and SN~1998S should be vanishingly small.
(Or that almost all the pre-existing
dust is evaporated at the moment of the UV flash).

For \el\ this result
is in contrast with the suggestion that the early IR excess is due 
to an IR echo caused by near-by dust.   

We suggest that the failure of the $\beta_{100,B}$ index
to display the presence of CSM around \el\ may be ascribed
to the assumption made to derive theoretically this index, namely the
assumption of spherically distributed dust.
 
The theoretical results of Emmering and Chevalier (1988)
on the IR echo light originated by an asymmetric CSM distribution around a SN
progenitor clearly illustrate how many parameters influence the observed results. 
Indeed they find that IR light curves can
substantially change by varying the geometrical properties of a
given non-spherical (but still axi-symmetric)
CSM distribution and/or its orientation relative to the line of sight. 
Even if not explicitly accounted for in their work
it can be nevertheless deduced that additional parameters, such as
a non-radial density profile of the dust, play an additional pivotal role.

We intend in the future to provide a tentative model of 
the dust distribution around \el\ . 
Finally we note that both for
SN~1998S (Fassia et al. 2000) and \el\ the observations do not rule out
a scenario of dust formation after the explosion. In the case of \el\
an occurrence like this would imply that the signature of dust appears both
at early and also at late stages.
In particular, for \el\, assuming A$_{V}$ $= 1.18$ mag,
the dereddened $K$ magnitude on JD+691 (see Table~\ref{nir:phot})
yields an observed flux of $1.2\times 10^{-13}$ erg cm$^{-2}$ s$^{-1}$ $\mu$m$^{-1}$.
As in Sect.~\ref{lc:d}, we can determine the radius of a
blackbody-emitting surface for a temperature of 1500 K, which is roughly
the sublimation temperature for dust grains. This amounts to $\sim 3 \times
10^{15}$ cm, of the order of the largest photospheric radius attained
(see Table~\ref{lboltab}),
a distance that might be covered by dust
formed in the ejecta moving with a velocity $\sim 1600$ km s$^{-1}$.
 
Therefore the scenario we propose embraces the evolution of the 
progenitor star in that cold, high-luminosity, mass loss
from evolved massive stars may proceed in episodes with asymmetric distributions.
It is however conceivable that single mass-loss episodes
are not completely stochastic in duration, periodicity and intensity.
Over long temporal intervals, possibly covering several
cycles of activity, it is possible to define
mean wind properties, such as, for example,
the dependence of the mean mass-loss rate on the main structural parameters such as
temperature and luminosity.
If the SN explosion occurs during a long pause in mass-loss we observe
a normal Type II SN, otherwise the typical features of the Type IIn
SNe will appear with varying degrees of visibility.
 
\section{Conclusions} 
Optical and NIR light curves have been presented 
for the Type IIn \el.  These data, in
comparison with those of SN~1998S,  suggest that the progenitor
star of \el\   was still undergoing a probably asymmetric mass loss episode 
at the moment of the explosion. 
Dust was already present in the expanding circumstellar envelope.
Dust lying in the
inner $\sim 10^{16}$ cm evaporated at the outburst.

Later, on, the evolution of NIR colors seems to be 
affected by the echo light from a pre-existing asymmetrically 
expanding envelope.  
In the absence of mass loss \el\ would have been presumably observed as a bona-fide
linear Type II SN. The way in which mass loss occurs 
in the last evolutionary phases of massive stars 
seems to be responsible for various
observed departures of Type II SNe from the standard II-L and II-P
behavior.

\begin{acknowledgements} 
 
This paper is partially based on observations made with the Italian Telescopio 
Nazionale Galileo (TNG) operated on the island of La Palma by the Centro  
Galileo Galilei of the CNAA (Consorzio Nazionale per l'Astronomia e  
l'Astrofisica) at the Spanish Observatorio del Roque de los Muchachos of the  
Instituto de Astrofisica de Canarias. The infrared data collected with the AZT-24 are products
of the Supernova Watch-dogging InfraRed Telescope (SWIRT), a joint project of the Astronomical
Observatories of Collurania-Teramo (Italy), Pulkovo (Russia) and Rome (Italy).
We thank F.\ Ghinassi and F.\ Mannucci for 
acquiring the latest NIR images with NICS.  We warmly thank Bruce Roscherr for advice 
and suggestions about the echo light issue.

\end{acknowledgements}

\end{document}